\documentclass[a4paper,11pt]{revtex4}
\usepackage{amsmath}
\usepackage{amssymb}
\usepackage{graphicx}
\usepackage[hypertex]{hyperref} % for hyperlinks between label-ref bibitem-cite
\def\da{{\partial a}}
\def\di{{\partial i}}
\begin{document}

\title{Decimation flows in constraint satisfaction problems}
\author{Saburo Higuchi$^{1,2}$ and Marc M\'{e}zard${}^{1}$}
\affiliation{
${}^1$Laboratoire de Physique Th\'eorique et Mod\`eles Statistiques,
CNRS and Universit\'e Paris-Sud, B\^{a}t 100, 91405 Orsay Cedex\\
${}^2$Department of Applied Mathematics and Informatics,
Ryukoku University, Otsu, Shiga, 520-2194, Japan}

\begin{abstract}
We study hard constraint satisfaction problems with a decimation approach based on message passing algorithms. Decimation induces a renormalization flow in the space of problems, and we exploit the fact that this flow transforms some of the constraints into linear constraints over GF(2). In particular, when the flow hits the subspace of linear problems, one can stop decimation and use Gaussian elimination. We introduce a  new decimation algorithm which uses this linear structure and shows a strongly improved performance 
with respect to the usual decimation methods on some of the hardest locked occupation problems.
\end{abstract}

\maketitle

\section{Introduction}
In recent years, statistical physics ideas and methods have become very useful in a number of problems involving many interacting variables in different branches of science.  A particularly fruitful area which is attracting a lot of attention is the one of constraint satisfaction problems. These are problems which appear in various areas of computer science and discrete mathematics. They include very fundamental problems like the satisfiability of random Boolean formulae \cite{cook1971ctp,cook1997fhi} or the coloring of graphs, and can also range all the way to very practical issues like chip testing or scheduling. Using 
statistical-physics-based approaches, the phase diagram of random ensembles of problems has been computed\cite{MPZ,mezard2002rks}, and new algorithms for finding solutions have been found\cite{SP}. There is now a whole field at the intersection of information theory, discrete mathematics and statistical physics\cite{mezard2009ipa}. 

The algorithms which have been so successful in satisfiability and coloring typically involve two main ingredients. The first one is a message passing method.
 It starts from a probability distribution on the set of interacting variables, which is typically the uniform measure on the configurations which satisfy all constraints. Then it aims at estimating, through mean-field based methods, the marginal distribution of each of the variables with respect to this distribution. 
 The second ingredient allows to make use of the information on marginals provided by the message passing. In the decimation process \cite{mezard2002rks,krzakala2007gsa}, one identifies, thanks to the messages, a strongly polarized variable and fixes it. The problem is thus reduced and the whole process (message passing and variable fixing) is iterated. The iteration of this process, when successful, eventually fixes all variables and finds an assignment which satisfies all constraints. Instead of decimation, one can also use a reinforcement procedure\cite{chavas2005spd}, but in the present paper we shall keep to decimation methods.

The decimation can be seen as a flow in the space of problems, where each step of the flow transforms an $N$-variable problem into a new $(N-1)$-variable problem. The choice of a heuristic to decide what variable to fix at each step defines a flow. The simplest choice is to use the belief propagation (BP) message passing procedure and fix the most polarized variable. This leads to a deterministic flow. An alternative choice, which results in a stochastic flow, consists in choosing at random one variable among the $R$ most polarized variables and fixing it. Belief propagation equations are  nothing but the self-consistency equations of the cavity method within a replica symmetric
approach; in some particularly difficult problems, when the multiplicity of metastable states is important, it is useful to go to more sophisticated replica-symmetry-broken cavity approach. This leads to the use of survey propagation (SP) as message passing, instead of  BP.  SP-based decimation and reinforcement are presently the best solvers for random satisfiability and coloring close to their SAT-UNSAT transition.
 Another useful improvement that we have found recently, which is efficient in some categories of hard problems, is to perform a special decimation where, instead of fixing one variable, one fixes the relative value between two 
variables\cite{HM1}. This requires to extend the message passing approach in order to 
obtain correlations between the variables, a technique which has been developed recently by several groups\cite{montanari2005clc,rizzokappen2007,MezMora,HM1,ktanaka2003,wellingyee2004}. 
The net result is also a decimation flow.

In this paper we propose a new use of decimation flows. Instead of trying to decimate the problem all the way until one has fixed all the variables and found a solution, we shall try to
 set up a decimation process
that brings the problem towards a subspace of problems that are not hard to solve.
Our approach is based on the remark that there exists a natural class of constraint satisfaction problems between binary variables which can be solved in polynomial time. 
These are the problems which can be written as systems of linear equations in GF(2). They can be solved by Gaussian elimination, which, in the worst case, takes a time growing like $N^3$. We shall call these problems `linear problems' (the reader should not be confused by the term 'linear': a linear problem does not necessarily have a linear energy function).
We shall thus seek a decimation flow which arrives in the subspace of linear problems, and then use Gaussian elimination. This strategy turns out to be rather powerful for a class of constraint satisfaction problems which are particularly hard to solve with usual methods, and in particular with the standard message passing + decimation strategy. These problems form a subclass of the locked occupation problems (LOPs),  introduced recently in\cite{ZM,ZM2}. It turns out that, for a whole class of these problems, the reason why they are hard to solve within message-passing/decimation is that they tend to flow towards linear problems, where variables (and even variable pairs) are typically unpolarized. This  makes the next decimation steps rather difficult. If one instead interrupts the decimation when the subspace of linear problems is reached by the decimation flow, the resulting algorithm shows much better performance.

The paper is organized as follows. Sec.\ref{sec:locked} introduces the locked occupation problems. Sec.\ref{sec:decflow} explains the decimation flows that we study; it describes the space of problems (weighted occupation problems) which are stable under this flow, and it defines the special type of constraints that can be encountered in decimation, in particular the linear constraints. In Sec.\ref{sec:locate} we explain how to identify the optimal variables (or pairs of variables) on which one  performs the decimation steps. Sec.\ref{se:numerics} summarizes the algorithm and shows the results of some numerical experiments. A short conclusion is given in Sec.\ref{se:conc}

\section{Locked Occupation Problems}
\label{sec:locked}
Locked occupation problems contain
$|V|=N$ binary variables taking values $x_i\in \{0,1\}$ $(i\in V)$. These variables are related by 
$|F|=M$ constraints, denoted by $a\in F$, each one  involving $k$ variables. To define these constraints, it is convenient 
to define a variable $x_i$ as `occupied' when $x_i=1$, and `empty' if $x_i=0$. Each constraint $a$ relates 
$k$ variables, with indices $i(a,1),\dots,i(a,k)$. It imposes some restrictions on the total
number of occupied variables among these $k$ variables. These restrictions  are fully   specified by a $(k+1)$-component `constraint-vector' $A$ with binary entries,
$A(r)\in\{0,1\}$. The constraint $a$ is satisfied if and only if 
\begin{equation}
  A\left(\sum_{m=1}^k x_{i(a,m)}\right)=1\ .
\label{eq:constraint}
\end{equation}

A factor graph $G=(V,F;E)$ is associated with every  instance of  a LOP in the usual way\cite{kschischang2001fga}.
The set of vertices of this bipartite graph $G$ is 
$V \cup F$ while the set of edges is $E=\{(i,a)| i\in V,a\in F, \exists j \;\text{s.t.}\; i=i(a,j)\}$. 
The notion of neighborhood is naturally introduced: 
$\partial a=\{i\in F|(i,a)\in E\}$,
$\partial i=\{a\in V|(i,a)\in E\}$.
For a collection of variables in $S\subset V$, 
we shall write $\underline{x}_S=\{x_i| i \in S\}$. We also use the short-hand notation $\underline{x}=\underline{x}_V$. 
The uniform measure over satisfying configurations can thus be written as 
\begin{equation}
P(\underline{x})=\frac{1}{Z}\prod_{a=1}^M\psi_a(\underline{x}_{\partial a})
\label{basic_measure}
\end{equation}
where  $\psi_a(\underline{x}_{\partial a})= A(\sum_{i\in \partial a} x_i)$.  It exists as soon as there is at least one configuration satisfying all constraints.

An occupation problem is \textsl{locked} if the following three conditions are met\cite{ZM,ZM2,zdeborova2008sph}.
\begin{itemize}
\item $A(0)=A(k)=0$.
\item $A(r)A(r+1)=0$ for $r=0,\ldots,k-1$.
\item Each variable appears in at least two constraints.
\end{itemize}

For example, the problem defined by
$k=3,\  A=(A(0),A(1),A(2),A(3))=(0,1,0,0)$ corresponds to
positive 1-in-3 satisfiability\cite{raymond2007pds}, 
while the problem with $k=4,\  A=(0,1,0,1,0)$ is a system of odd-parity checks.

\section{Decimation flow}
\label{sec:decflow}

In this paper we concentrate on finding a satisfying assignment with
large probability, assuming 
that there is at least one such assignment.
To this end, 
we make use of `flow' operations which replace an instance $I$ of the constraint satisfaction problem at hand
with another instance $I'$ which has less variables. 
We will focus on two basic flow steps, which consist in either fixing a variable to a given value, or fixing the relation between two variables.
Our procedure thus generalizes the usual decimation by introducing also the possibility of `pair-fixing'. This strategy follows from 
the observation in \cite{HM1} that the correlations between variables play an important role in LOPs. 
The first case (single-variable decimation) consists in imposing 
\begin{equation}
x_{i}=x\label{eq:deg1constraint}
\end{equation}
where $ x$ is $0$ or $1$.
The second one (pair decimation)  consists in imposing 
\begin{equation}
x_{i}= x_j+y \bmod 2 \label{eq:deg2linearconstraint}
\end{equation}
where $ y$ is $0$ or $1$.
In the next section we will explain with what kind of approximate methods one can identify a variable $x_i$ and its assigned value $x$ for single-variable decimation, or  
a pair $x_i,x_j$ and its assigned relative value $y$ for pair decimation. Here we want to study the flow process itself. 

First we notice that each step effectively reduces the
number of variables by one. At each decimation we update a table which contains all the variables which have been fixed, and to what value (either a fixed value or a relation to another variable). When the decimation is completed, this table allows to find the value of all variables. As usual in renormalisation group flows, our basic flow steps, when applied to a LOP,
produce a problem (instance) which is no longer a LOP. One thus needs to identify a space of problems, containing the LOPs,  which is stable under our two basic decimation steps.
Such a stable space is provided by the {\it weighted} occupation problems.

\subsection{Weighted occupation problems}
We generalize the occupation problems as follows. We first allow the constraint degree $k$ and constraint-vector $A$ to be dependent on the constraint: the constraint $a$ involves $k_a$ variables and its constraint-vector can depend on $a$  and is denoted as $A_a$.
Each constraint $a$ depends furthermore on $k_a$ integer weights $w_{a,i}$, $i\in\da$, and on a shift $s_a$.
In a weighted occupation problem, the constraint $a$ is satisfied if and only if the variables $x_i,i\in\da$ are such that:
\begin{equation}
A_a\left(  \sum_{i\in\da} w_{a,i} x_i+s_a\right) =1 \ .
\label{eq:constraint-local}
\end{equation}

Let $W$ be the set of weighted occupation problems.
We shall now show that $W$ is stable under our two basic decimation steps, and make explicit  the two flow steps in $W$ which correspond  to the two decimation procedures.

\subsection{Flow equations}
Let us first study  a single variable decimation,  fixing $x_i$ to a value $x\in\{0,1\}$. The variable $x_i$ disappears from the problem.
The constraints $a$ with $a\in\di$ are modified: their degree is decreased by $1$, as they no longer involve variable $i$, and their 
constraint vector is shifted by a constant:
\begin{equation}
  s'_a=s_a+w_{a,i}x \quad \text{for} \; a\in\partial i. 
\end{equation}

We now turn to a pair decimation operation. Suppose that we fix 
$x_i=x_j+y\ \bmod 2$. Explicitly, this amounts to the operation on integers:
\begin{equation}
x_i=x_j+y(1-2 x_j)
\end{equation}
Then the variable $x_i$ disappears from the problem, the number of variables is reduced by one, all the edges $(i,a)$ for $a\in\di$ disappear from the factor graph. The constraints $a\in\di$ are modified,
but their modification depends on whether $j\in\da$ or not. 

If a constraint $a\in\di$ does not involve $j$, then $j$ is added to the neighborhood of $a$. The modification of $a$ is given by:
\begin{equation}
  E'=E\cup (j,a)\ \ ; \ \ (\partial a)'=(\partial a) \cup (j)\ \  ;\ \ 
  w'_{a,j}=w_{a,i} \; (1-2y) \ \ ; \ \ 
  s'_{a}=s_a+w_{a,i} \; y\ .
\end{equation}
Notice that the degree $k_a$ of $a$ is unchanged in this case: through decimation, $a$ loses one neighbor ($i$), and gains another one ($j$).

If a constraint $a$ involves both $i$ and $j$, then:
\begin{equation}
  w'_{a,j}=w_{a,j}+(1-2y)\; w_{a,i}\ \ ; \ \ 
  s'_{a}=s_a+w_{a,i}\;  y\ .
\end{equation}
In this procedure, the degree of $a$ is decreased by one (if $ w'_{a,j} \neq 0$), or by two (if $ w'_{a,j} =0$, in which case both $i$ and $j$ disappear from $\da$).

Clearly, $W$ is stable through our decimation operations. It is important to notice that, at each step:
\begin{itemize}
\item The number of variables decreases
\item The degree $k_a$ of each constraint is non-increasing.
\item The function $A_a$ is unchanged.
\item The sum $\sum_{i\in\partial a} |w_{a,i}|$ is non-increasing for each constraint.
\end{itemize}
The flow generated through these operations can be seen as a kind of  renormalization group flow.
If we start from a LOP where each constraint $a$ has degree $k_a=k$, this flow stays within a finite subspace of $W$, where $k_a\le k$ and 
$\sum_{i\in\partial a} |w_{a,i}|+s_a\le k$. These properties allow to use this type of flow as an effective algorithm.

\subsection{Special constraints}
When applying the flow process, one may encounter  constraints which present some peculiarities which should be noticed.

{\it Irrelevant variables}. First of all it may happen that a variable  is irrelevant for some  constraint. With some appropriate labelling of the variables, consider the constraint $A(\sum_{i=1}^kw_ix_i+s)=1$.
Variable $1$ is irrelevant if and only if:
\begin{equation}
\forall \{x_2,\dots,x_k\}\in\{0,1\}^{k-1}:\ \ A\left(\sum_{i=2}^k w_ix_i+s\right)=A\left(w_1+\sum_{i=2}^k w_ix_i+s\right)
\end{equation}
When a variable is irrelevant for a constraint, it can just be eliminated from this constraint, reducing $k$ by one, without any other change in the weights or threshold. We shall suppose here that, whenever a constraint is changed by the flow, one checks for possible irrelevant variables and eliminates them.

{\it Linear constraints}.  Let us consider a constraint $A(\sum_{i=1}^kw_ix_i+s)=1$ which has no irrelevant variable. This constraint is a linear constraint if and only if the set ${\mathcal B}$ of configurations of the $k$ variables $x_1,\dots,x_k$ which satisfy this constraint can be characterized by an affine subspace in GF(2). This is the case whenever there exists $k+1$ numbers $a_1,\dots,a_k,b\in\{0,1\}$ such that ${\mathcal B}$ consists of all solutions of the equation   $\sum_ {i=1}^ka_ix_i+b=0\; \pmod 2$. A typical case where this happens is when $w_i=1$ and the vector $A$  is either given by $\forall\; r:\; A(r)=\frac12(1-(-1)^r)$ or by $\forall\; r:\;  A(r)=\frac12(1+(-1)^r)$. But it may happen that a constraint is linear in a less obvious way. For instance consider the constraint on $k=2$ variables characterized by 
 $w_1=-2$,$w_2=3$, $s=2$ and the vector $A=(110001)$. This is a linear constraint (which can be rewritten as $x_1+x_2=1 \; \pmod 2$).

{\it Variable polarizing constraints}. 
Consider the constraint $A(\sum_{i=1}^kw_ix_i+s)=1$. This constraint is said to be polarizing for the variable $x_1$ if and only if there exists $\tau\in\{0,1\}$ such that
\begin{equation}
\forall \{x_2,\dots,x_k\}\in\{0,1\}^{k-1}:\ \ A\left(\sum_{i=2}^kw_ix_i+s\right)=\tau=1-A\left(w_1+\sum_{i=2}^kw_ix_i+s\right)
\end{equation}
If $\tau=1$, the constraint imposes that $x_1$ should be equal to $1$. If $\tau=0$, it imposes that $x_1$ should be equal to $0$.

{\it Pair polarizing constraints}. 
Consider the constraint $A(\sum_{i=1}^kw_ix_i+s)=1$. This constraint is said to be polarizing for the pair $x_1,x_2$ if and only if there exists $\tau\in\{0,1\}$ such that
\begin{equation}
\forall \{x_3,\dots,x_k\}\in\{0,1\}^{k-2}:\ \left\{\begin{array}{l}
 A\left(\sum_{i=3}^kw_ix_i+s\right)=A\left(w_1+w_2+\sum_{i=3}^kw_ix_i+s\right)=\tau\\
 A\left(w_1+\sum_{i=3}^kw_ix_i+s\right)=A\left(w_2+\sum_{i=3}^kw_ix_i+s\right)=1-\tau
\end{array}
\right.
\end{equation}
If $\tau=1$, the constraint imposes that $x_1+x_2=0\; \pmod 2$. If $\tau=0$, it imposes that  $x_1+x_2=0\; \pmod 2$.

\section{How to find  polarized variables and correlated pairs}
\label{sec:locate}
In this section we explain how to choose a variable, or a pair of variables, to which the flow operations are applied.

One first natural class is found when the flow itself generates variable- or pair- polarizing constraints as defined in the previous section. As soon as the flow generates a constraint, it can be tested with respect to these two criteria, and if the constraint is found to be variable- or pair- polarizing, one performs the corresponding variable- or pair- decimation. Notice that the variable decimation process induced in such a way is equivalent to the well known ``warning propagation'' procedure \cite{mezard2002rks,montanari2007scs}, which is 
 frequently used  with BP and SP in order to infer the consequences of fixing a variable.

Next we consider the subset of all the linear constraints defined in the previous section. This subset obviously constitutes a necessary condition for the constraint satisfaction problem at hand. That condition can in principle imply relations of the form 
(\ref{eq:deg1constraint}) or (\ref{eq:deg2linearconstraint}).
If the standard Gaussian elimination on GF(2) is applied on the subset of linear constraints, all the variable that can be solved in the form (\ref{eq:deg1constraint}) are found.
Some of the two variable relations (\ref{eq:deg2linearconstraint}) can also be found even if $x_i$ and $x_j$ are not neighbors.
The set of pairs found in this way depends on the detailed order of operations in the Gaussian elimination.
If more than one such variables or pairs are found, one performs variable- or pair- decimation.
This step does more than the standard warning propagation. It can find some of distant but completely correlated pairs in a single step. How much portion of them are found depends on the nature of instances.

When the above procedures do not impose any strict relations on variables or pairs, one must estimate the single variable marginals
$P_i(x)=P(x_i=x)$ and the relative pair marginals $P_{ij}(y)=P(x_i+x_j=y)$. 
The standard methods for finding polarized variables are  BP\cite{yedidia2003ubp} and SP\cite{mezard2002rks} .
In fact, application of single-variable decimation defined above on a variable which has extreme 1-variable marginal according to BP or SP is equivalent to BP or SP-guided decimation \cite{montanari2007scs}. In this case, one  stays  in a subspace where all the 
 weights on the edges are equal to one.

To find a highly correlated pair, one should resort to some of the recently proposed methods which aim at estimating correlations with message passing \cite{montanari2005clc,rizzokappen2007,MezMora,HM1,ktanaka2003,wellingyee2004}.
The application of the second flow operation within the  susceptibility propagation method \cite{MezMora,HM1}  is equivalent to pair decimation\cite{HM1}.

The degree of polarization of a variable $i$ is conveniently measured by the entropy of the marginal, $S_i=-\sum_x P_i(x)\log P_i(x)$. Similarly, we define the relative pair entropy as $S_{ij}= -\sum_y P_{ij}(y)\log P_{ij}(y)$. Our primary aim is thus to find a strongly polarized (low entropy) variable or pair, as in \cite{HM1}. On top of this primary goal, one can also decide to favor the decimation steps which lead to a larger fraction of linear constraints. In the most general case within our setting, one just decides what operation to do (variable- or pair- fixation) by looking for variables (resp. pairs) with small $S_i$ (resp. $S_{ij}$) such that the corresponding fixation induces some linear constraints. Many types of penalty functions can be used, putting more or less weight on the variables, the pairs, or the induction of linear constraints. In the next section we shall show that the results obtained with some of the simplest scheme is already much better than all existing methods, on some classes of hard LOPs.

\section{Numerical Experiments}
\label{se:numerics}
\subsection{The algorithm}
\label{se:alg}
In our experiments we have used a simple version of the general class of algorithms defined above. In this version, when we generate a new constraint, we check whether it is a variable-polarizing constraints, and if this is the case we implement the corresponding variable decimation step. If the constraint has degree $2$, we also check whether it is a pair-polarizing constraint, and if this is the case we implement the corresponding pair-decimation step. Then we check for the linear constraints and run Gaussian elimination on them. (Notice that these steps can also be seen as a simple `on-line' exploitation of the linear structure of the problem). In all other situations we run belief propagation, estimate the corresponding variable entropies $S_i$, and fix the variable $i$ with the smallest entropy. When all the constraints are linear, we stop the decimation process and
apply a Gaussian elimination procedure on the corresponding linear problem. When the decimation leads to a problem with a small enough number of variables, we perform an exhaustive search. The structure of the program is thus the following:

\texttt{Input:}
 Factor graph, constraint-vector, convergence criterion, initial messages

\texttt{Output:}
A satisfying assignment (or FAIL-NOT-FOUND)

\begin{itemize}

\item While instance is either with more than $D$ variables or equivalent to a linear binary equation:
   \begin{itemize}
      \item Compute local entropy estimates for the 1-variable marginals.
      \item Apply single-variable decimation to the variable with the smallest entropy. Return FAIL-NOT-FOUND if 
there is a contradiction.
      \item Apply recursively the following steps until no further change occurs:
      \begin{itemize}
         \item Eliminate irrelevant variables
         \item Identify  variable-polarizing constraints and perform the corresponding variable-fixing step.  Return FAIL-NOT-FOUND if there is a contradiction.
         \item Among the degree 2 constraints, identify pair-polarizing constraints and perform the corresponding pair-decimation step. Return NOT-FOUND if there is a contradiction.
         \item Identify the subset of all the linear constraints and run Gaussian elimination on them. Perform the corresponding variable- and pair- fixing step.  Return FAIL-NOT-FOUND if there is a contradiction.
      \end{itemize}
  \end{itemize}
  \item If the whole set of constraints is a linear binary system on GF(2), perform Gaussian elimination. If the linear problem admits a solution,
   \begin{itemize}
         \item Then return the satisfying assignment
         \item Else return FAIL-NOT-FOUND
   \end{itemize}
   \item When the number of variables is equal to or smaller than $D$:  perform an exhaustive search for satisfying assignments. If found
   \begin{itemize}
          \item Then return the satisfying assignment
          \item Else return FAIL-NOT-FOUND
   \end{itemize}
\end{itemize}

\subsection{Class of problems}
We have studied various LOPs defined from random factor graphs, uniformly chosen among the graphs with the following degrees. All function nodes have degree $k$ and the variables have random degrees chosen from truncated Poisson degree distribution
\begin{equation}
  q(\ell)=
  \begin{cases}
    0 & (\ell=0,1)\\
    \frac{\mathrm{e}^{-c}c^\ell}{\ell! (1-(1+c)\mathrm{e}^{-c})}. & (\ell\geq2)
  \end{cases},
\end{equation}
for which the average degree is 
\begin{equation}
  \overline{\ell}=\sum_{\ell=0}^\infty \ell q(\ell)=\frac{c(1-\mathrm{e}^{-c})}{(1-(1+c)\mathrm{e}^{-c})}.
\end{equation}
The number of function nodes $M$ and the number of variables $N$ are related by $M k = N \overline \ell$. We are interested in the thermodynamic limit where $N$ and $M$ go to infinity at fixed $k, \overline \ell$.

The phase diagram of these problems has been studied in \cite{ZM,ZM2}. Like most random constraint satisfaction problems, they exhibit  two main phase transitions when the average degree $ \overline{\ell}$  increases (this increases  the density of constraints). For $\overline \ell<\ell_d$ the problem is typically easy, for $\ell_d<\overline\ell<\ell_s$ there exists with probability one (in the large $N$ limit) a solution, but the space of solutions is made of isolated configurations which are very far away from each other. For $\overline\ell>\ell_s$ the problem has no solution with probability one. In problems like 1-or-3-in-5 SAT (described by the vector $A=010100$)   or 1-or-4-in-5 SAT (described by the vector $A=010010$), the usual BP+decimation or BP+reinforcement algorithms are unable to find solutions
 when they are applied to  instances in the intermediate phase $\ell_d<\overline \ell <\ell_s$. We have thus focused on this region.

\subsection{Results}
\begin{figure}
  \centering
  \includegraphics[scale=0.8]{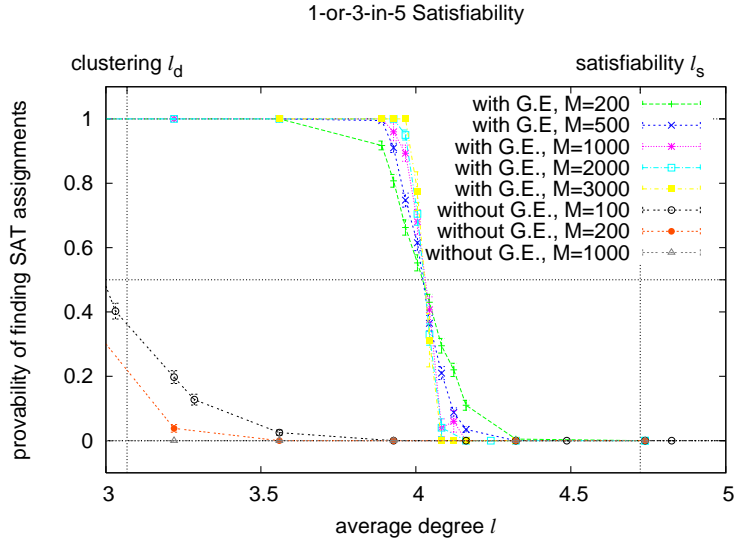}
  \caption{Probability to find a solution for 1-or-3-in-5 satisfiability, plotted versus the average degree $\overline \ell$. The clustering threshold $\ell_d$ and the satisfiability threshold $\ell_s$ are shown by vertical lines. The results from our new algorithms are signalled by 'with G.E.' and show a clear threshold behaviour around $\overline \ell=4$. For comparison we show the result obtained with the standard  BP decimation (curves `without G.E.'), which is unable to find solutions when   $\overline \ell>\ell_d=3.07$ (in fact its threshold is well below the clustering point $\ell_d$).
In both algorithms we use 3 restarts.}
\label{fig:zbe-1or3in5}
\end{figure}

The LOP
1-or-3-in-5 satisfiability is a 
choice candidate for exploiting the idea of flow towards a linear system, because if a variable connected to a given factor is fixed to $0$, the corresponding factor is transformed into a 1-or-3-in-4 constraint, which is a linear node.
Figure~\ref{fig:zbe-1or3in5} show the result of experiments for this problem, with the algorithm of Sect.\ref{se:alg} used with $D=16$.

It can be observed that the algorithm works well in the lower half $\overline{\ell}\precsim 4.0$ of the clustered phase.
By inspecting the decimation process in this lower half closely, we find that all the instances flow to a  linear problem, consisting only of 1-or-3-in-4 constraints, in approximately $M$ steps. This means that, in the early stage, a variable in the neighborhood of each constraint is fixed to 0. Then the Gaussian elimination finds some solutions. (Notice that, at this stage the problem is linear and  all the 1-variable marginal are unpolarized\cite{mezard2009ipa}; therefore, if instead of 
using Gaussian elimination one would keep on with BP, the next decimation steps would be done totally at random).
On the other hand, on the upper half $\overline{\ell}\succsim4.0$, 
the algorithm  also flows to a linear problem but it turns out that this problem has no solution.
When $\overline{\ell}<\ell_s$,  we know a priori that the instances all have satisfiable solutions with probability one. This means that the algorithm has made a fatal error in guessing in the early stage. 

While the new algorithms is still not able to reach the satisfiability threshold, it widely increases the set of instances for which one can find a solution, with respect to  existing algorithms.
We have observed  that this improvement also occurs also in the LOPs
1-or-3-in-6 and 2-or-4-in-6.

\begin{figure}
  \centering
  \includegraphics[scale=0.8]{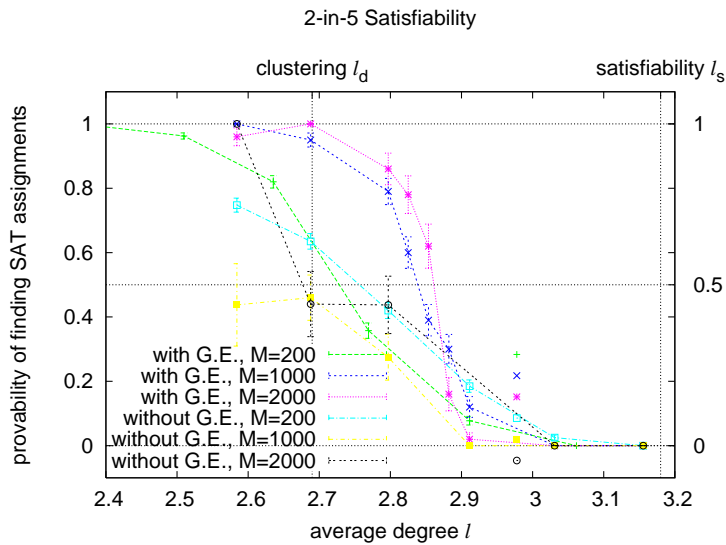}
  \caption{Probability to find a solution  for 2-in-5 satisfiability, same conditions as in Fig.\ref{fig:zbe-1or3in5}}
\label{fig:zbe-2in5}
\end{figure}

On the other hand, the improvement is much less impressive for the following problems: 1-in-3, 1-in-4, 1-in-5, 1-in-6, 2-in-4, 2-in-5, 2-in-6, 3-in-6,1-or-4-in-5, 1-or-5-in-6. 
For instance, Fig.~\ref{fig:zbe-2in5} shows the performance for 2-in-5. The reason for this relatively poorer performance is that theses problems typically do not flow to XORSAT. The improvement with respect to standard decimation is thus only due to the `on-line' use of linear constraints when one identifies pair-polarizing or linear constraints and performs the subsequent decimation.

\section{Conclusions}
\label{se:conc}
The random LOPs are typically hard to solve in their clustered phase. An exception is given by linear problems, where all the constraints can be written as linear equations over GF(2), which can be solved in polynomial time by Gaussian elimination. In this paper we have shown how one can exploit this peculiarity within the decimation approach. The idea is that, instead of trying to follow the decimation flow all the way until all variables are fixed, one stops the flow whenever it hits the subspace of linear problems, and then uses Gaussian elimination. This idea is also complemented by the use of Gaussian elimination `on-line' through the detection of polarizing constraints and linear constraints.
This strategy has been found to greatly improve the range of problems that can be solved efficiently in the case of 1-or-3-in-5, 
1-or-3-in-6 and 2-or-4-in-6 LOPs. In other problems the improvement is mainly due to the on-line elimination ans is less impressive, but still present.

The present approach was the simplest implementation of this idea, and one could think of several directions in which to develop it. 
We have  done a few experiments  in which we tried to accelerate the flow towards the subspace of linear problems  by favouring the decimation steps which lead to more linear constraints. We have seen that  this approach typically improves the efficiency, but it is not evident that it improves the threshold in the $N\rightarrow+\infty$ limit. More work along this direction needs to be done in order to find the optimal protocol and its threshold.

\section*{Acknowledgment}
S.H. was supported by Ryukoku University Research Fellowship (2008).

% generated by bibtex
\bibliographystyle{apsrev}  
\bibliography{bp}

\end{document}